\documentclass[12pt]{JHEP3}

\usepackage{amsmath,amssymb,graphicx,epsf}

\newcommand{\tr}{\hbox{tr}}

\newcommand{\ZZ}{\mathbb{Z}}
\newcommand{\ID}{\mathbb{I}}

\title{A note on the center of mass in non-commutative theories}

\author{Robert C. Helling\\

School of Engineering and Science\\
International University Bremen\\
Postfach 750561\\
28725 Bremen, Germany\\

{\tt E-mail: helling@atdotde.de}}
\preprint{IUB-TH-053 \\hep-th/0503104}

\abstract{The dynamics of a stack of $N$ D-branes is described by
  $U(N)$ gauge theory of which the central $U(1)$ describes the center
  of mass motion and the remaining $SU(N)$ describes the internal
  dynamics. In the non-commutative situation, these two parts are
  coupled by the $*$-commutator interaction. We describe here how to
  identify the correct decoupled U(1) center of mass subsector of 
  $U(N)$ gauge theory for the case of the non-commutative torus. The
  internal dynamics remainder is {\em not} a theory of $SU(N)$
  valued fields but has a simple description in momentum space. }

\begin{document}


\section{Introduction}
\setcounter{equation}{0}

The low energy theory of a  stack of $N$ D-branes is simply a $U(N)$
gauge theory with a number of scalar fields $X^i$ in the adjoint
representation. The diagonal elements of these scalar field matrices
are usually interpreted as giving the position of the individual
branes of the stack in the transversal direction while the
off-diagonal elements arise from strings stretching between different
branes. 

For a flat target space, the set-up is translationally invariant in
the transversal directions. Thus, by conservation of momentum, it
should be possible to separate off the free centre of mass motion of
the system from the internal dynamics. This is realized by observing
that the central $U(1)\subset U(N)$ drops out of all commutators that
make up the interaction and indeed the centre of mass coordinates
$\frac 1N\tr X^i$ are free. The remaining internal dynamics is then
given by a $SU(N)$ gauge theory.

In the presence of background fields, the world-volume of the D-branes
is known to become a non-commutative space\cite{DH, CDS, S, SW}. This
results in the commutators being replaced by $*$-commutators
\begin{equation*}
  [f,g]_* = f*g-g*f=\frac 12(f^a*g^b+g^b*f^a)[T^a,T^b]+ \frac
  12(f^a*g^b-g^b*f^a)\{T^a,T^b\} 
\end{equation*}
in terms of a basis of the representation of the gauge group. The
second term however poses a problem as the anti-commutator of two
representation matrices in general is not a representation
matrix. Especially in the case of $SU(N)$, the anti-commutator in
general is not trace-less and thus the internal dynamics $SU(N)$ seems
to mix with the centre of mass motion given by the trace\cite{Armoni}.

There is a similar problem of defining non-commutative gauge theories
for other groups that are not just products of $U(N_i)$ factors. In
those cases, the problem however is not as pressing as it is not in
conflict with physical problems as the decoupling of the centre of mass
motion. The later follows from  translation invariance (note that the
background fields inducing the non-commutativity are tangent to the
branes while we are discussing transversal translations which are not
broken by the presence of the background fields): Especially the brane
realization of $SO$ and $Sp$ gauge theories requires orientifold
branes that project out the Kalb-Ramond field responsible for the
non-commutativity and exceptional gauge groups are even more difficult
to obtain from brane physics in smooth targets.

In the past, there have been several attempts to solve this problem at
hand: Notably, there is \cite{JSSW} using enveloping algebras and the
Seiberg-Witten map. This treatment is quite formal and it remains to
be shown if the problem does not reappear once the enveloping algebra
is represented on the fields. Furthermore, doing perturbation theory
in the non-commutativity parameter bears the possibility of treating a
non-local theory as a local one thereby obscuring important global
properties (for example the crucial relation $U^q=V^q=1$ below is
invisible in finite order perturbation theory in $\theta$).

The case of $SO$ and $Sp$ non-commutative gauge theories is also
considered in \cite{BSST}. That treatment however relies on a
$B_{\mu\nu}$-field that is discontinuous at the location of the
D-branes. Finally, there is also \cite{CD} which uses open Wilson
lines to define a non-commutative $SU(N)$ gauge theory.

\section{The non-commutative $U(1)_{CM}$}
As a $B$-field in the macroscopic directions would violate Lorentz
invariance, we will only consider non-commutativity in the compact
directions. For concreteness, we will discuss a stack of $N$ D2-branes
that wrap a non-commutative torus $T_\theta$. We will use the same
notation as \cite{GHLL} and as explained there, we will make again use
of the fact that by compactification we can integrate the dimension-full
two-form $B$ to a dimensionless parameter $\theta$.

\EPSFIGURE{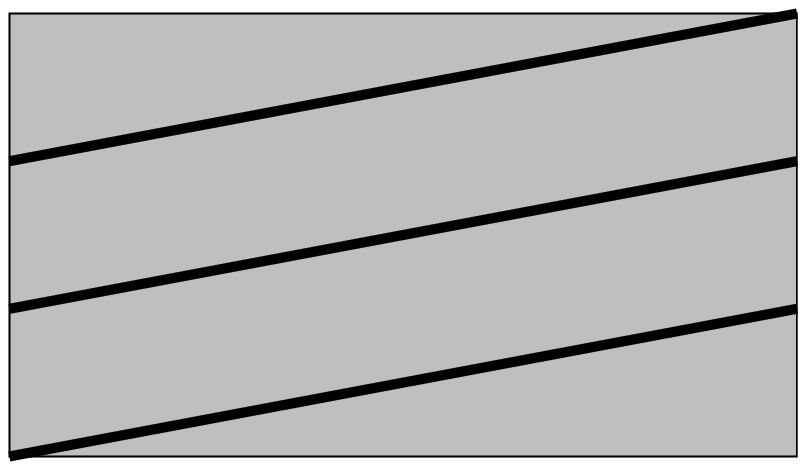,width=0.3\hsize}{D1-brane dual to $\theta=1/3$}
Under T-duality in one direction of the torus, the D2-branes becomes
flat D1-branes that wrap the torus. In order to avoid the marginal
case of D1's that ergodically wrap the volume of the T-dual torus, we
further restrict ourselves to the case of rational $\theta=p/q$. This
corresponds to D1's that wraps $p$ times one cycle of the torus while
wrapping the other cycle $q$ times.

Next, we expand all fields into their Fourier modes (we generically
write $f,g,\ldots$ for fields like $A_\mu$ and $X^i$ in the adjoint of
$U(N)$ and suppress all commutative coordinates):
\begin{equation*}
  f(x_1,x_2)=\sum_{m_1,m_2}f_{m_1,m_2}e^{2\pi im_1x_1}e^{2\pi
    im_2x_2}=\sum_{m_1,m_2}f_{m_1,m_2}U^{m_1}V^{m_2}, 
\end{equation*}
where we introduced $U=\exp(2\pi i x_1)$ and $V=\exp(2\pi i x_2)$. The
modes $f_{m_1,m_2}$ take values in the Lie algebra of $U(N)$. The
non-commutativity is summarized by
\begin{equation*}
  UV=e^{-2\pi i p/q}VU.
\end{equation*}
Our ordering convention is that all $U$'s are
left of the $V$'s. The $*$-product thus reads
\begin{equation*}
  (f*g)_{m_1,m_2}=\sum_{n_1,n_2}f_{n_1,(m_2-n_2)}g_{(m_1-n_1),n_2}
  e^{2\pi i (m_2-n_2)(m_1-n_1)p/q}.  
\end{equation*}
Furthermore, it is useful to split all the mode labels modulo $q$ as
$m_1=qM_1+\mu_1$ and similar with the understanding that Greek labels
take values in $\ZZ_q$ as the phase factor only depends on the later:
\begin{equation*}
  (f*g)_{m_1,m_2}=\!\!\!\!\sum_{N_1,\nu_1,N_2,\nu_2}\!\!\!f_{N_1q+\nu_1,
    (M_2q+\mu_2-N_2q-\nu_2)} g_{(M_1q+\mu_1-N_1q-\nu_1),N_2q+\nu_2}  
  e^{2\pi i (\mu_2-\nu_2)(\mu_1-\nu_1)p/q}.  
\end{equation*}
Now our aim is to split all the fields $f$ into a free (abelian,
commutative) $\tilde f$ that describes the center of mass motion and a
decoupled $\hat f=f-\tilde f$ that describes the internal dynamics. In
the commutative situation, this is simply achieved by defining $\tilde
f$ to be the $U(1)$ part $\frac 1N\tr (f)$ of $f$ while the traceless $\hat f$
is in the remaining $SU(N)$.  As explained in the introduction, this
is not sufficient in the non-commutative situation as the commutator
interaction mixes these two parts.

To find the appropriate decoupled $U(1)$, it is essential to realise
that the elements $U^q$ and $V^q$ are central, that is, they commute
with all functions on the non-commutative torus. Thus we are led to
define $\tilde f$ to contain the traces of the $(\mu_1,\mu_2)=(0,0)$
parts only:
\begin{equation*}
  \tilde f(x_1,x_2)=\sum_{M_1,M_2}\ID\frac 1N\tr f_{qM_1,qM2}U^{qM_1}V^{qM_2},
\end{equation*}
where $\ID$ is the unit matrix in the Lie algebra of $U(N)$. As above,
we take $\hat f=f-\tilde f$. Then it is easy to check that indeed
$\tilde f$ decouples, that is
\begin{equation}
  \label{crucial}
  [\tilde f,g]_*=0 \qquad\hbox{and}\qquad \widetilde{[f,g]_*}=0.
\end{equation}
The theory of the $\tilde f$'s is indeed a commutative $U(1)$ gauge
theory that we interpret as to describe the centre of mass motion of
the stack of branes.

Note that our definition is non-local in the following sense: Instead
of taking $SU(N)$ and $U(1)$ valued matrices over each point we only
generalise $SU(N)$- and $U(1)$-valued functions to the non-commutative
case by making use of the Fourier decomposition. Therefore we modify
the gauge group rather than the structure group.
But this is in the
spirit of non-commutative geometry where attention is shifted from
individual points to the algebra of functions.

Furthermore, the second equation in (\ref{crucial}) suggests the
generalization of $SU(N)$-valued functions to more general
non-commutative spaces (starting with the torus with irrational
$\theta$): These are the functions that can be written as linear
combinations of $*$-commutators of $U(N)$-valued functions\footnote{We
  thank P. Schupp for suggesting this}. There are, however, thorny
analytical problems to be solved: Probably, one should consider the
closure of the image of the commutator (e.g. allow infinite,
converging sums) but the case of the irrational torus shows that one
must not include too many limit points if one does not want the fine,
ergodic structure of that theory. We leave this for future
investigation. Similarly, we hope to gain better understanding of the
relation of our definition of $\hat f$ with the approach of
\cite{JSSW}, that defines the  non-commutative $SU(N)$ theory in terms
of the image of the Seiberg-Witten map of its commutative
counter-part. A counting of degrees of freedom however shows that this
relation, if existent, has to be non-local.

Let us close this note with two related comments: First, as explained
in \cite{GHLL}, it is a manifestation of Morita equivalence that the
algebra of $U$ and $V$ can be realized in terms of clock an shift
matrices in $U(q)$. In that description, we are dealing with a
$U(N)\otimes U(q)$ gauge theory on a commutative torus with both radii
smaller by a factor $q$. Our split $f=\tilde f+\hat f$ there
corresponds to the split of that gauge group into $U(1)$ and
$S(U(N)\otimes U(q))$. This interpretation also implies that the
internal theory of the $\hat f$ can be defined as a quantum theory
as it can be rewritten as an ordinary commutative gauge theory.

Second, the fact that only every $q$th Fourier mode appears in $\tilde
f$ can be motivated as well in the T-dual description: There the
wrapping stack of D1-branes looks like $q$ D1-stacks. Thus only every
$q$th mode describes the collective motion of the $q$ stacks which is
the center of mass motion, see the figure.

\goodbreak
\acknowledgments\nobreak The author is grateful for support by a grant of the
DFG Schwerpunkt Stringtheorie and for discussions with Peter Schupp
and Dieter L\"ust.

\bibliographystyle{JHEP}
\bibliography{nccm}

\providecommand{\href}[2]{#2}\begingroup\raggedright\begin{thebibliography}{1}

\bibitem{DH}
M.~R. Douglas and C.~M. Hull, {\it D-branes and the noncommutative torus},
  {\em JHEP} {\bf 02} (1998) 008,
  [\href{http://xxx.lanl.gov/abs/hep-th/9711165}{{\tt hep-th/9711165}}].

\bibitem{CDS}
A.~Connes, M.~R. Douglas, and A.~Schwarz, {\it Noncommutative geometry and
  matrix theory: Compactification on tori},  {\em JHEP} {\bf 02} (1998) 003,
  [\href{http://xxx.lanl.gov/abs/hep-th/9711162}{{\tt hep-th/9711162}}].

\bibitem{S}
V.~Schomerus, {\it D-branes and deformation quantization},  {\em JHEP} {\bf 06}
  (1999) 030, [\href{http://xxx.lanl.gov/abs/hep-th/9903205}{{\tt
  hep-th/9903205}}].

\bibitem{SW}
N.~Seiberg and E.~Witten, {\it String theory and noncommutative geometry},
  {\em JHEP} {\bf 09} (1999) 032,
  [\href{http://xxx.lanl.gov/abs/hep-th/9908142}{{\tt hep-th/9908142}}].

\bibitem{Armoni}
A.~Armoni, {\it Comments on perturbative dynamics of non-commutative yang-
  mills theory},  {\em Nucl. Phys.} {\bf B593} (2001) 229--242,
  [\href{http://xxx.lanl.gov/abs/hep-th/0005208}{{\tt hep-th/0005208}}].

\bibitem{JSSW}
B.~Jurco, S.~Schraml, P.~Schupp, and J.~Wess, {\it Enveloping algebra valued
  gauge transformations for non- abelian gauge groups on non-commutative
  spaces},  {\em Eur. Phys. J.} {\bf C17} (2000) 521--526,
  [\href{http://xxx.lanl.gov/abs/hep-th/0006246}{{\tt hep-th/0006246}}].

\bibitem{BSST}
L.~Bonora, M.~Schnabl, M.~M. Sheikh-Jabbari, and A.~Tomasiello, {\it
  Noncommutative so(n) and sp(n) gauge theories},  {\em Nucl. Phys.} {\bf B589}
  (2000) 461--474, [\href{http://xxx.lanl.gov/abs/hep-th/0006091}{{\tt
  hep-th/0006091}}].

\bibitem{CD}
C.-S. Chu and H.~Dorn, {\it Noncommutative su(n) and gauge invariant baryon
  operator},  {\em Phys. Lett.} {\bf B524} (2002) 389--394,
  [\href{http://xxx.lanl.gov/abs/hep-th/0110147}{{\tt hep-th/0110147}}].

\bibitem{GHLL}
Z.~Guralnik, R.~C. Helling, K.~Landsteiner, and E.~Lopez, {\it Perturbative
  instabilities on the non-commutative torus, morita duality and twisted
  boundary conditions},  {\em JHEP} {\bf 05} (2002) 025,
  [\href{http://xxx.lanl.gov/abs/hep-th/0204037}{{\tt hep-th/0204037}}].

\end{thebibliography}\endgroup

\end{document}